# Electrical Degradation of Vertical β-$Ga_2O_3$ Schottky Barrier Diodes Under Au-Ion Irradiation to 1.3×$10^7$ $cm^{-2}$

Botong Li[1], Imteaz Rahaman[1], Bingcheng Da[2], Houqiang Fu[2], Michael A. Scarpulla[1,3,*], Kai Fu[1,*]

[1]*Department of Electrical and Computer Engineering, University of Utah, Salt Lake City, UT 84112, USA*

[2]*School of Electrical, Computer, Energy Engineering, Arizona State University, Tempe, Arizona 85281, USA*

[3]*Department of Materials Science and Engineering, University of Utah, Salt Lake City, UT 84112, USA*

[*]Corresponding author: mike.scarpulla@utah.edu; kai.fu@utah.edu

**Abstract:**

The electrical degradation and time-dependent current of vertical β-$Ga_2O_3$ Schottky barrier diodes (SBDs) subjected to energetic Au-ion irradiation were systematically investigated. Two irradiation exposures were performed under a constant reverse bias of −100 V to evaluate the effects of heavy-ion irradiation on device performance and current behavior. Following irradiation, the devices exhibited degraded forward conduction, increased specific on-resistance, enhanced reverse leakage current, and a reduction in apparent carrier concentration. During irradiation, pronounced current increases and transient current spikes were observed. After each beam termination, the leakage current exhibited an initial rapid decrease followed by a slower process. Single-exponential fitting revealed distinct fast and slow components, while a double-exponential model provided a more comprehensive description of the full decay behavior. Following the second irradiation exposure, the fast and slow characteristic time constants increased from 33.50 to 40.64 s and from 190.35 to 282.29 s, respectively, indicating prolonged recovery dynamics under repeated irradiation. A charge-trapping model is proposed to explain the irradiation-induced current enhancement and subsequent fast and slow relaxation through modulation and gradual recovery of the effective Schottky tunneling barrier. These findings provide insight into the relationship between heavy-ion-induced electrical degradation and time-dependent current in β-$Ga_2O_3$ SBDs.

Ultra-wide-bandgap (UWBG) semiconductors have attracted considerable attention for next-generation power electronics due to their superior material properties and potential for high-voltage and high-power operation[1,2]. Among these materials, β-$Ga_2O_3$ has emerged as a promising candidate due to its large bandgap of approximately 4.85 eV, high critical electric field of ~8 MV/cm, and the availability of large-area single-crystal substrates[3–5]. These advantages provide opportunities for developing high-performance power devices with improved voltage-blocking capabilities and reduced conduction losses. In particular, β-$Ga_2O_3$ Schottky barrier diodes (SBDs) have been extensively investigated for high-voltage rectification and power conversion applications[6–9].

The ability to maintain stable electrical performance under irradiation is essential for power electronic devices used for aerospace, satellite, and other radiation-exposed applications[10]. In such environments, energetic particles can interact with semiconductor materials, generating electron–hole pairs, introducing lattice defects, and modifying the electrical characteristics of devices[11–14]. These interactions can lead to increased leakage current, carrier removal, enhanced on-resistance, and even permanent device failure[15,16]. The large bandgap and high threshold displacement energy of β-$Ga_2O_3$ make it an attractive material for high-voltage electronics in harsh environments[17,18]. However, these intrinsic material advantages do not necessarily guarantee radiation tolerance, since the electrical response of a device also depends strongly on irradiation-induced defects, carrier trapping, and the resulting changes in carrier transport mechanisms. Therefore, understanding the performance of β-$Ga_2O_3$ devices under irradiation is essential for evaluating their reliability for applications.

Previous studies have investigated the radiation response of β-$Ga_2O_3$ devices under different irradiation conditions. Particular attention has been given to single-event effects (SEEs)[19,20], such as transient leakage-current responses and single-event burnout (SEB)[21,22], which can cause severe degradation or catastrophic failure under high electric fields. These investigations have provided important insights into the vulnerability of β-$Ga_2O_3$ power devices to energetic particle strikes. In addition to these single-event phenomena, irradiation can introduce electrically active defects and trapped charges that progressively modify the device characteristics during repeated exposure. Such defects may affect carrier concentration, Schottky contact behavior, and reverse leakage current by introducing additional carrier transport pathways[23–25]. In particular, the accumulation of irradiation-induced defects and trapped charges may influence both the magnitude of leakage current and its behavior after the ion beam is removed.

In this work, we have systematically investigated the real-time electrical performance of vertical β-$Ga_2O_3$ SBDs subjected to Au-ion irradiation exposures under a constant reverse bias of −100 V. The current decay behavior is observed after beam termination, exhibiting distinct fast and slow components. Single- and double-exponential fitting models are applied to quantify the characteristic times and their evolution after successive irradiation cycles. The changes in device performance are also characterized through current–voltage (I–V) and capacitance–voltage (C–V) measurements before and after irradiation, with particular emphasis on forward conduction, specific on-resistance, reverse leakage current, and the carrier concentration. An interface-charge-trapping model is proposed in which trapped charge modulates the effective Schottky barrier height, accounting for the current increase and its recovery process. These results provide insight into the relationship between heavy-ion-induced damage and defect-related carrier transport in β-$Ga_2O_3$ SBDs.

Vertical β-$Ga_2O_3$ SBDs were fabricated on a (001)-oriented β-$Ga_2O_3$ wafer consisting of $n^+$ Sn-doped β-$Ga_2O_3$ substrate and a 10-μm-thick n-type epitaxial drift layer (Novel Crystal Technology, Inc., Japan). Figure 1(a) illustrates the device structure and fabrication sequence. Mesa patterns were defined using a 7-μm-thick AZ10XT photoresist mask, followed by inductively coupled plasma etching using $BCl_3$/Ar chemistry with RF and ICP powers of 40 and 200 W, respectively. The resulting circular mesas had a diameter of 320 μm and an etch depth of approximately 3.2 μm. Circular Schottky anodes with a diameter of 300 μm were subsequently patterned by photolithography. A Ni/Au/Ni/Au stack with thicknesses of 50/100/50/130 nm was deposited and defined by lift-off. A Ti/Au stack with thicknesses of 50/100 nm was deposited on the backside to form the cathode. After fabrication, the β-$Ga_2O_3$ sample was cleaved and mounted on a test fixture, using Al wire bonding for connections, as shown in Figs. 1(b)–(d).

Au-ion irradiation was performed using the K500 cyclotron at Texas A&M University Cyclotron Institute. The Au beam energy was 15.0 MeV/u. The devices were irradiated and tested in air at normal incidence. Two successive irradiation exposures were performed on the same device in this work. The first exposure delivered a fluence of $1.0 \times 10^7$ $cm^{-2}$ over a recorded beam live time of 698.9 s, with an average flux of $\sim 1.4 \times 10^4$ $cm^{-2} \cdot s^{-1}$. The second exposure delivered an additional fluence of 2.8 × $10^6$ $cm^{-2}$, with the similar flux. The cumulative fluence after both exposures was about 1.3 × $10^7$ $cm^{-2}$. During the irradiation test, the diode was biased at −100 V, and its reverse current was continuously recorded before, during, and after beam exposure. The current compliance was set to 100 nA on the measurement instrument.

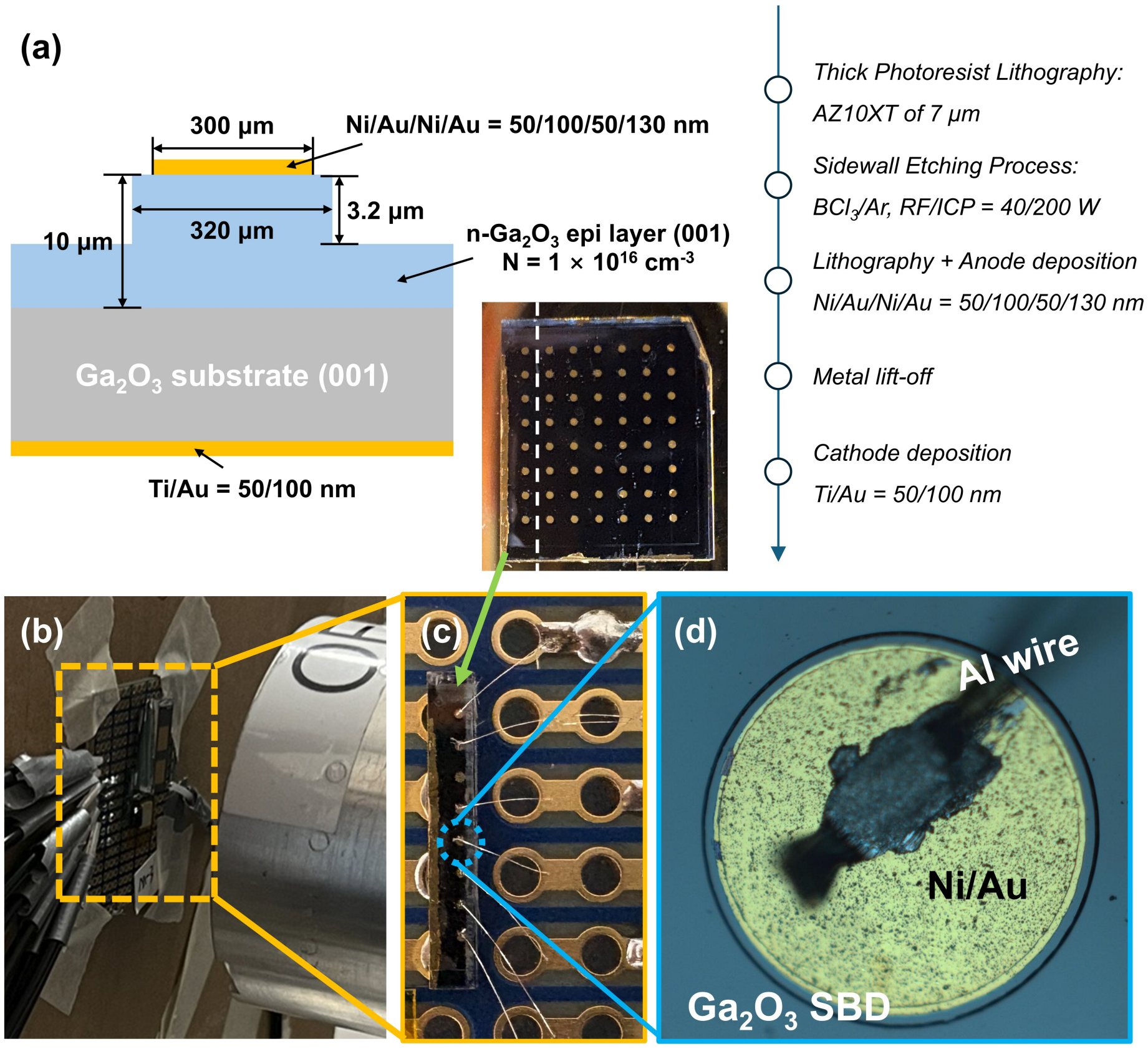

**FIG. 1.** (a) Schematic and fabrication process of β-$Ga_2O_3$ SBD used in radiation tests. (b) Photo of radiation test setup. (c)-(d) Photos of the β-$Ga_2O_3$ SBDs.

Figure 2(a) presents the linear energy transfer (LET) profile of the incident Au ions in β-$Ga_2O_3$ as a function of penetration depth. Although the LET gradually increases with depth, its variation within the depletion region at a reverse bias of −100 V is relatively small, allowing the LET to be treated as approximately constant at ~66.6 MeV/(mg/$cm^2$) in this region. The calculated projected ion range is 55.8 μm, substantially exceeding the epitaxial drift-layer thickness. Therefore, the incident Au ions are expected to traverse the drift layer while depositing energy along their trajectories rather than stopping within it.

The forward current density–voltage (J–V) characteristics were measured before irradiation and 24 h after irradiation, and plotted on linear and semi-logarithmic scales in Figs. 2(b) and 2(c), respectively. Following irradiation, the turn-on voltage increases slightly from approximately 0.9 to 1.0 V, accompanied by a reduction in forward current density. The extracted ideality factor increases from 1.04 to 1.08, whereas the apparent Schottky barrier height changes only slightly, from 1.15 to 1.14 eV. These results indicate a modest degradation in forward conduction and an increased deviation from ideal thermionic-emission behavior. Fig. 2(d) further shows that the minimum extracted specific on-resistance increases from 29.8 to 32.7 $m\Omega \cdot cm^2$. Figure 2(e) also shows an increase in current hysteresis between the voltage sweeps after irradiation. Figure 2(f) compares the reverse J–V characteristics before and after irradiation. A pronounced increase in reverse leakage current is observed following irradiation. To investigate the possible leakage mechanism, the post-irradiation characteristics were fitted using trap-assisted tunneling (TAT), Poole–Frenkel (PF) emission, and variable-range hopping (VRH) models. The TAT model provides the highest coefficient of determination, with $R^2 = 0.9897$, compared with 0.9644 for PF emission and 0.9588 for VRH. Within the bias range investigated, this better agreement supports a contribution from trap-assisted tunneling to the enhanced leakage current. The result is consistent with irradiation-induced defect states providing additional pathways for carrier transport. Figure 2(g) presents the capacitance–voltage (C–V) characteristics and the corresponding apparent carrier concentration profiles before and after irradiation. A slight decrease in the extracted carrier concentration is observed after irradiation, as seen in the inset of Fig. 2(g). Integration of the C–V curves over the bias interval from 0 to −8 V, normalized by the device area and elementary charge, yields a sheet charge density of $3.27 \times 10^{11}$ $cm^{-2}$ before irradiation and $2.76 \times 10^{11}$ $cm^{-2}$ after irradiation. This corresponds to a reduction of approximately 15.6% in the charge modulated over this bias interval. Together with the carrier concentration profiles, this reduction is consistent with partial carrier removal or compensation associated with irradiation-induced defects. The changes in specific on-resistance, ideality factor, and apparent Schottky barrier height due to radiation are summarized in Fig. 2(h).

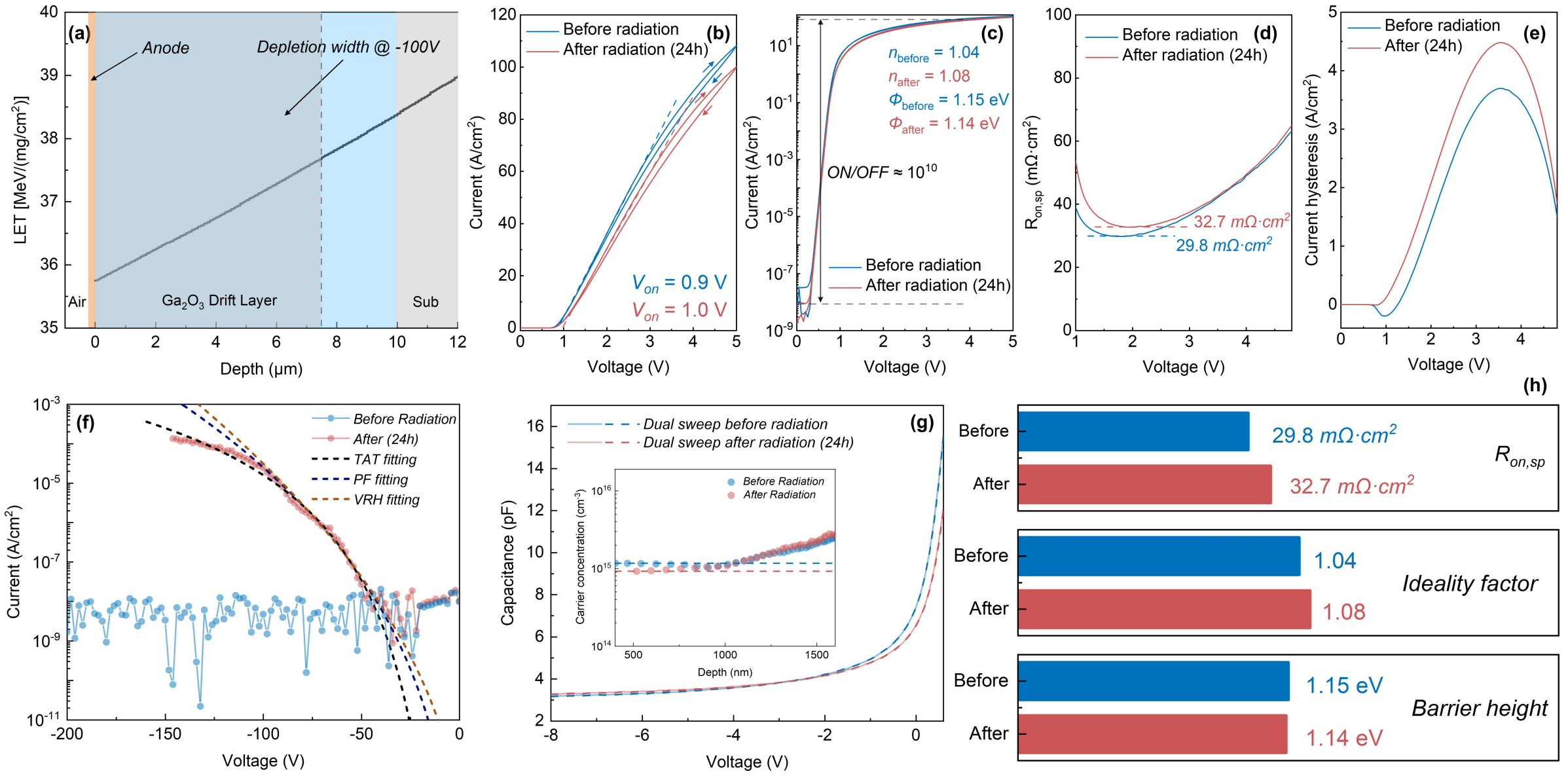

**FIG. 2.** (a) LET profile of incident Au ions as a function of penetration depth. Forward J–V characteristics before and after Au irradiation in (b) linear scale and (c) semi-log scale. (d) Specific on-resistance vs. voltage before and after irradiation. (e) Current hysteresis as a function of voltage before and after irradiation. (f) Reverse J–V curves before and after irradiation along with the TAT, PF, and VRH fits. (g) C–V characteristics measured using dual voltage sweeps before and after irradiation; inset: apparent carrier concentration vs. depth. (h) Comparison of the extracted specific on-resistance, ideality factor, and Schottky barrier height before (blue) and after irradiation (red).

Figure 3 presents the time-dependent reverse leakage current of the β-$Ga_2O_3$ SBD under a constant bias of −100 V during two Au-ion irradiation cycles, plotted on linear and semi-logarithmic scales, with the sampling step of 0.1 s. Based on the flux and the anode area, approximately 10 Au ions struck the SBD per second during irradiation. Upon the first beam exposure, the current increases abruptly from its pre-irradiation level, consistent with the generation and collection of electron–hole pairs during electronic stopping of the Au ions. During beam ON, six fast-decaying transient current spikes and a larger current excursion toward the end of the first exposure are both observed, suggesting single-event-induced current responses. After the first beam exposure is terminated, the current initially decreases rapidly and subsequently exhibits slower relaxation, approaching its pre-irradiation level after approximately 1200 s. This slow decay can be attributed to the gradual release of charge trapped at irradiation-generated defects, as discussed below. During the second irradiation cycle, substantially larger current excursions are observed, and the measured current reaches the preset compliance limit of 100 nA, prompting termination of the beam exposure. After the beam is off, the current exhibits an initial rapid decrease followed by a prolonged decay. In contrast to the first cycle, the leakage current remains substantially above its original level throughout the subsequent observation period. This incomplete recovery indicates permanent degradation of the β-$Ga_2O_3$ SBD, attributable to accumulated irradiation damage.

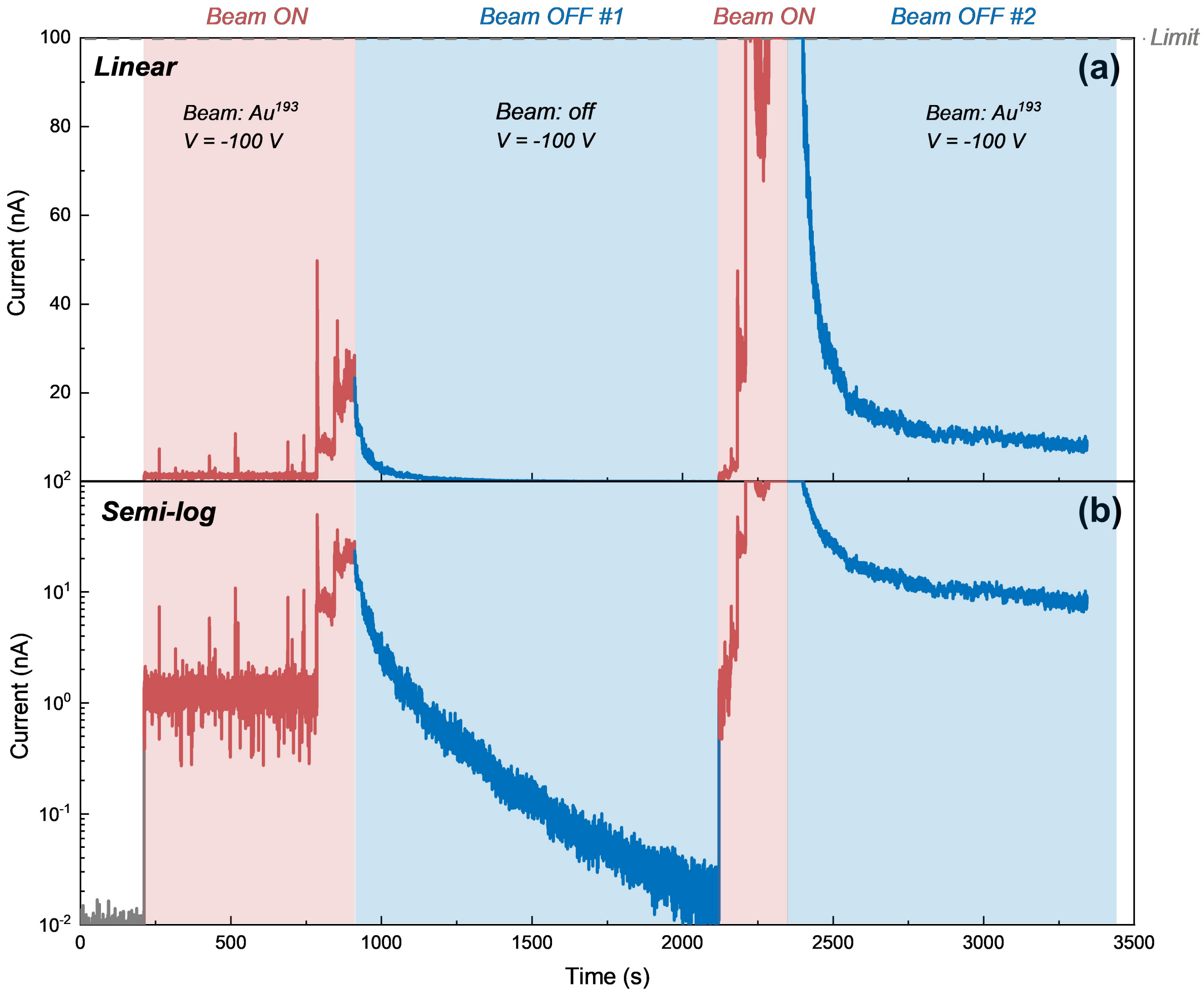

**FIG. 3.** Time-dependent current of a β-$Ga_2O_3$ Schottky barrier diode held at −100 V during two cycles of Au irradiation in (a) linear scale and (b) log scale.

To further investigate the post-irradiation behavior, the current transients were analyzed using exponential decay models, as shown in Figs. 4(a) and 4(b). The single-exponential model was first applied, expressed as[26,27]

$$I(t) = I_0 + Ae^{-t/\tau} \quad (1)$$

where $I_0$ represents the asymptotic current, A is the decay amplitude, and $\tau$ is the characteristic time. As shown in Fig. 4(a), a single-exponential function cannot adequately describe the entire first decay curve. However, fitting the initial and later portions separately yields good agreement within each region, with characteristic time constants of 33.50 and 190.35 s, respectively. These results reveal distinct fast and slow decay regimes, indicating that two relaxation processes with different time scales contribute. To account for contributions with different time scales, a dual-exponential model was also employed[28,29]:

$$I(t) = I_0 + A_1 e^{-t/\tau_1} + A_2 e^{-t/\tau_2} \quad (2)$$

The time constants $\tau_1$ and $\tau_2$ were fixed at the values obtained from the separate single-exponential fits. The resulting double-exponential curve closely reproduces the experimental data across the full decay interval as shown by the red line in Fig. 4(a), consistent with the coexistence of fast and slow contributions. The same procedure was also applied to the second beam-off interval, as shown in Fig. 4(b). Separate single-exponential fits yield fast and slow time constants of 40.64 and 282.29 s, respectively. As summarized in Fig. 4(c), both characteristic time constants increase after the second irradiation cycle, coinciding with the increase in accumulated Au-ion fluence.

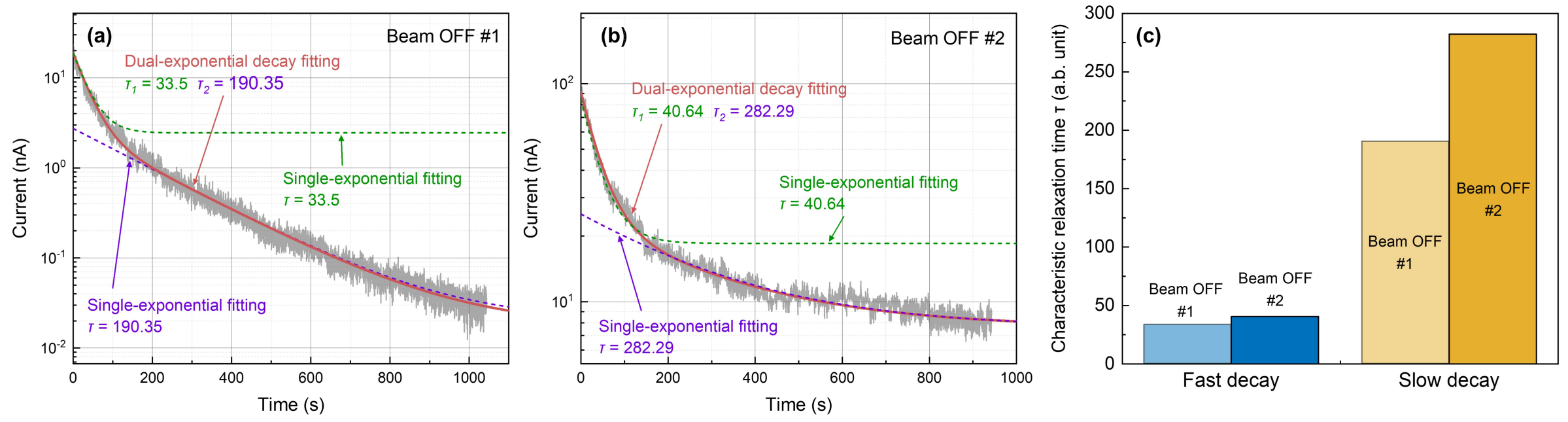

**FIG. 4.** Single-exponential and double-exponential fitting for (a) first Au-ion irradiation and (b) second Au-ion irradiation. (c) Comparison of characteristic time constants between first decay and second decay for single-exponential fitting.

Figure 5 illustrates a proposed mechanism for the irradiation-induced current increase and subsequent relaxation in the reverse-biased β-$Ga_2O_3$ SBD. Before irradiation, the Schottky junction suppresses electron injection from the metal into the semiconductor, resulting in a low reverse leakage current. Upon beam exposure, energy deposition by the incident Au ions generates electron–hole pairs, whose separation and collection under the applied electric field produce an immediate increase in current, as illustrated in Fig. 5(a). Figures 5(b) and 5(c) schematically represent hole trapping at two defect types, shallow $T_1$ and deep $T_2$, associated with the fast and slow relaxation components identified in Fig. 4. In the proposed model, net positive trapped charge near the junction enhances the local electric field and reduces the effective tunneling width of the Schottky barrier, thereby increasing electron injection under reverse bias. Changes in the populations or occupancies of these states can consequently produce additional current increases during irradiation. After beam termination, as shown in Fig. 5(d), direct generation of electron–hole pairs cease, while the trapped-charge distribution continues to evolve through carrier emission, recombination, and possible retrapping. As the excess trapped charge is neutralized or released, the associated field enhancement diminishes and the effective tunneling barrier gradually recovers, producing a decreasing leakage current. The simultaneous relaxation of rapidly and slowly recovering charge populations provides a physical interpretation of the double-exponential decay: the fast contribution diminishes predominantly during the initial recovery, whereas the slow contribution sustains the later current tail. During the second exposure, the defect and charge configuration remaining from the first irradiation can modify the response to additional energy deposition. Further trapping and defect generation may enhance local carrier injection and introduce additional trap-assisted tunneling pathways, as illustrated in Fig. 5(e). This interpretation is consistent with the stronger current response during the second irradiation and the improved agreement of the post-irradiation reverse characteristics with the trap-assisted tunneling model. Following beam termination, longer-lived trapped charge and persistent transport pathways may contribute to the prolonged decay and elevated residual leakage depicted in Fig. 5(e). This framework links the transient current relaxation with the post-irradiation electrical degradation, while the microscopic defect identities and their individual contributions remain unresolved. Figure 5(f) summarizes the corresponding current evolution using a simplified schematic of the measured I–t response, showing the current increase during each irradiation exposure, the subsequent beam-off decay, and the incomplete recovery following the second exposure.

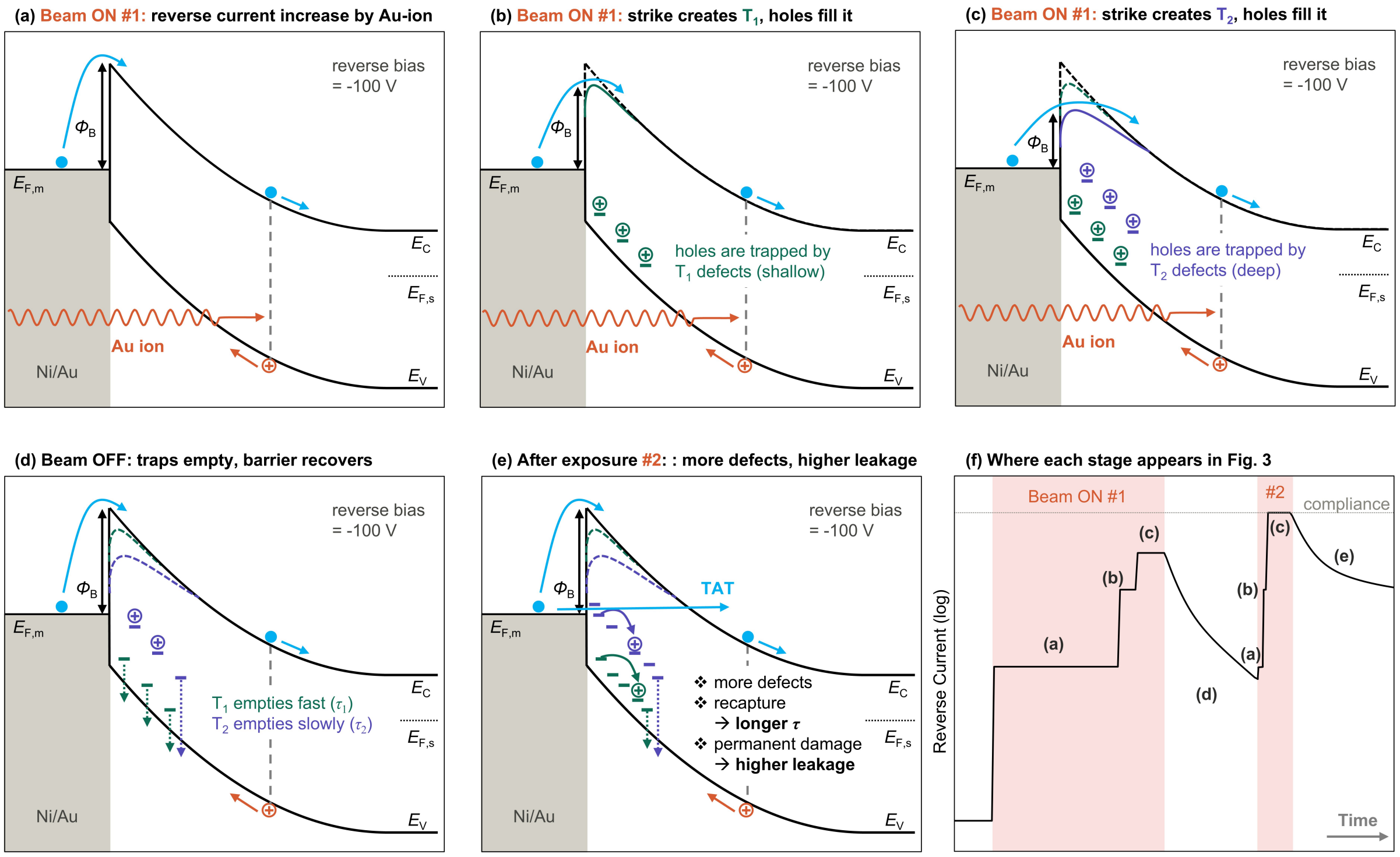

**FIG. 5.** Schematic of the proposed mechanism for radiation effects in the reverse-biased β-$Ga_2O_3$ SBD. (a) Increase of reverse leakage current due to the Au-ion beam. (b) Further increase of reverse leakage current due to holes trapped at $T_1$ defects (shallow). The positive charge of the trapped holes reduces the effective tunneling width of the Schottky barrier. (c) Further increase of reverse leakage current due to holes trapped at $T_2$ defects (deep). The positive charge of the trapped holes further reduces the effective tunneling width. (d) Decrease of reverse leakage current due to release of holes during Beam OFF. Two time constants correspond to the shallow and deep defects. (e) Decrease of reverse leakage current due to release of holes during Beam OFF after Beam ON #2, accompanied by additional defects, hole recapture, and permanent damage. (f) Schematic I–t curve indicating where each stage (a)–(e) occurs in Fig. 3.

In summary, the electrical degradation and time-dependent current of vertical β-$Ga_2O_3$ SBDs under Au-ion irradiation exposures were investigated. Following irradiation, the devices exhibited increased specific on-resistance, ideality factor, and reverse leakage current, along with a reduction in apparent carrier concentration, indicating degradation of electrical performance. The enhanced reverse leakage current was analyzed using different conduction models, with trap-assisted tunneling providing the best agreement with the experimental results, suggesting a contribution from irradiation-induced defect states. During irradiation, pronounced current increases and transient current spikes were observed, followed by a rapid initial decrease and slower decay after beam termination. The first irradiation cycle exhibited nearly complete current recovery, whereas an elevated leakage current remained after the second exposure, indicating persistent electrical degradation and/or long-lived trapped charge. Quantitative analysis using single- and double-exponential fitting revealed the coexistence of fast and slow components, with both characteristic time constants increasing after repeated irradiation. The proposed model attributes the current enhancement to trapped-charge-induced barrier thinning, the subsequent decay to charge release and neutralization, and the elevated residual leakage to long-lived trapped charge and persistent defect-assisted transport pathways. These results highlight the role of defect-related processes in the electrical degradation and post-irradiation current of β-$Ga_2O_3$ SBDs, providing insight into their reliability under energetic heavy-ion exposure.

## ACKNOWLEDGMENTS

The authors acknowledge the support from the University of Utah High Tech West (HTW) Award, startup fund, Partners for Innovation, Ventures, Outreach and Technology (PIVOT) Energy Accelerator under Grant U-7352FuEnergyAccelerator2023, and Research Incentive Seed Grant by the Price College of Engineering and the Vice President for Research (VPR) Office. This work made use of Nanofab EMSAL shared facilities of the Micron Technology Foundation Inc. Microscopy Suite sponsored by the John and Marcia Price College of Engineering, Health Sciences Center, Office of the Vice President for Research.

## AUTHOR DECLARATIONS

### Conflict of Interest

The authors have no conflicts to disclose.

### Author Contributions

**Botong Li:** Conceptualization (equal); Data curation (lead); Formal analysis (equal); Investigation (equal); Methodology (equal); Visualization (equal); Writing - original draft (lead); Writing - review & editing (equal). **Imteaz Rahaman:** Formal analysis (supporting); Investigation (supporting); Methodology (supporting); Writing - review & editing (supporting). **Bingcheng Da:** Methodology (supporting); Investigation (supporting). **Houqiang Fu:** Methodology (supporting); Resources (supporting). **Michael A. Scarpulla:** Conceptualization (equal); Investigation (supporting); Methodology (supporting); Project administration (lead); Resources (lead); Supervision (lead); Validation (equal); Writing – original draft (supporting); Writing – review & editing (lead). **Kai Fu:** Conceptualization (equal); Investigation (supporting); Methodology (lead); Project administration (lead); Resources (lead); Supervision (lead); Validation (equal); Writing – original draft (equal); Writing – review & editing (lead).

## DATA AVAILABILITY

The data that support the findings of this study are available from the corresponding authors upon reasonable request.

## REFERENCES

1. M.H. Wong, O. Bierwagen, R.J. Kaplar, and H. Umezawa, "Ultrawide-bandgap semiconductors: An overview," Journal of Materials Research 36(23), 4601–4615 (2021).

2. Y. Zhang, D. Dong, Q. Li, R. Zhang, F. Udrea, and H. Wang, "Wide-bandgap semiconductors and power electronics as pathways to carbon neutrality," Nat Rev Electr Eng 2(3), 155–172 (2025).

3. M. Higashiwaki, "β-Ga2O3 material properties, growth technologies, and devices: a review," AAPPS Bull. 32(1), 3 (2022).

4. C. Wang, J. Zhang, S. Xu, C. Zhang, Q. Feng, Y. Zhang, J. Ning, S. Zhao, H. Zhou, and Y. Hao, "Progress in state-of-the-art technologies of Ga2O3 devices," J. Phys. D: Appl. Phys. 54(24), 243001 (2021).

5. S. Yang, Z. Song, X. Tan, X. Liu, S. Wu, Z. Li, R. Yao, H. Yang, Y. Yan, Y. Wang, Z. Yu, Y. Zhang, X. Yi, J. Wang, and J. Li, "Heteroepitaxial Growth of Ga2O3 Film: Recent Progress and Perspectives," Mater. Adv., (2026).

6. Y. He, F. Zhao, B. Huang, T. Zhang, and H. Zhu, "A Review of β-Ga2O3 Power Diodes," Materials 17(8), 1870 (2024).

7. S. Sun, C. Wang, S. Alghamdi, H. Zhou, Y. Hao, and J. Zhang, "Recent Advanced Ultra-Wide Bandgap β-Ga2O3 Material and Device Technologies," Advanced Electronic Materials 11(1), 2300844 (2025).

8. L.-Q. Zhang, J.-J. Liu, Y.-T. Tian, H. Xi, Q.-H. Yue, H.-F. Li, Z.-Y. Wu, and L.-F. Sun, "Recent Progress of β-Ga2O3 Power Diodes: A Comprehensive Review," Inorganics 13(11), 364 (2025).

9. S. Bu, Y. Wang, X. Zheng, S. Yue, D. Lin, L. Yi, V. Melikyan, X. Ma, and Y. Hao, "Forward bias stress-induced degradation mechanism in β-Ga2O3 SBDs: A trap-centric perspective," Appl. Phys. Lett. 126(12), 122104 (2025).

10. A.G. Chmielewski, "Radiation technologies: The future is today," Radiation Physics and Chemistry 213, 111233 (2023).

11. V.A.P. Aguiar, S.G. Alberton, and M.S. Pereira, "Radiation-Induced Effects on Semiconductor Devices: A Brief Review on Single-Event Effects, Their Dynamics, and Reliability Impacts," Chips 4(1), 12 (2025).

12. A. Azarov, J.G. Fernández, J. Zhao, F. Djurabekova, H. He, R. He, Ø. Prytz, L. Vines, U. Bektas, P. Chekhonin, N. Klingner, G. Hlawacek, and A. Kuznetsov, "Universal radiation tolerant semiconductor," Nat Commun 14(1), 4855 (2023).

13. I. Filipev, J. Paino, J. Poder, D. Cutajar, N. Hardcastle, S. Guatelli, M. Petasecca, M. Lerch, V. Feygelman, T. Kron, and A. Rosenfeld, "Review on high spatial resolution dosimetry with pixelated semiconductor detectors for radiation therapy," Radiation Measurements 177, 107272 (2024).

14. D.M. Fleetwood, “Perspective on radiation effects in nanoscale metal–oxide–semiconductor devices,” Appl. Phys. Lett. 121(7), 070503 (2022).

15. N. Manikanthababu, H. Sheoran, P. Siddham, and R. Singh, “Review of Radiation-Induced Effects on β-Ga2O3 Materials and Devices,” Crystals 12(7), 1009 (2022).

16. “Contact property depending on radiation intensity between the perovskite semiconductor layer and electrode film | Applied Physics Letters | AIP Publishing,” (n.d.).

17. Z. Muhammad, Y. Wang, Y. Zhang, P. Vallobra, S. Peng, S. Yu, Z. Lv, H. Cheng, and W. Zhao, “Radiation-Tolerant Electronic Devices Using Wide Bandgap Semiconductors,” Advanced Materials Technologies 8(2), 2200539 (2023).

18. H.D. Ellis, A. Mallapillai, J. Miller, I. Rahaman, B. Li, V. Sarkar, and K. Fu, “Dosimetry for proton therapy using a β-Ga2O3 metal–semiconductor–metal detector with low-noise amplification,” Semicond. Sci. Technol. 41(4), 045014 (2026).

19. J. Gui, X. Song, B. Mei, Y. Sun, H. Lv, L. Yu, D. Li, G. Zhang, X. Sun, C. Cao, S. Zhao, Y. Hao, and J. Zhang, “Analysis of the single-event effects in β -Ga2O3 Schottky barrier diodes,” Applied Physics Letters 127(4), 042103 (2025).

20. X. Xia, J.-S. Li, R. Sharma, F. Ren, M.A.J. Rasel, S. Stepanoff, N. Al-Mamun, A. Haque, D.E. Wolfe, S. Modak, L. Chernyak, M.E. Law, A. Khachatrian, and S.J. Pearton, “Radiation Damage in the Ultra-Wide Bandgap Semiconductor Ga2O3,” ECS J. Solid State Sci. Technol. 11(9), 095001 (2022).

21. X. Song, S. Zhao, J. Gui, N. Wang, K. Wang, L. Du, Z. Kong, L. Yu, G. Zhang, L. Gao, S. You, Y. Hao, and J. Zhang, “Single-event burnout mechanism in β -Ga2O3 Schottky barrier diodes,” Applied Physics Letters 127(18), 182111 (2025).

22. S. Yue, X. Zheng, F. Zhang, R. Zhang, Y. Li, T. Zhu, Y. Wang, X. Ma, and Y. Hao, “Study on the Single Event Burnout Mechanism of β-$Ga_2O_3$ Schottky Barrier Diode Under Heavy Ion Irradiation,” IEEE Transactions on Electron Devices 71(12), 7377–7382 (2024).

23. J.F. McGlone, H. Ghadi, E. Cornuelle, A. Armstrong, G. Burns, Z. Feng, A.F.M.A. Uddin Bhuiyan, H. Zhao, A.R. Arehart, and S.A. Ringel, "Proton radiation effects on electronic defect states in MOCVD-grown (010) β-Ga2O3," J. Appl. Phys. 133(4), 045702 (2023).

24. Z. Zhang, X. Lin, T. Wang, C. Liu, J. Liu, L. Shu, Y. Sun, M. Huo, and L. Xiao, "Trap-Induced Leakage Current Increase in β-Ga2O3 Schottky Barrier Diodes Under 473-MeV Kr Ion Irradiation," IEEE Transactions on Nuclear Science 72(11), 3543–3550 (2025).

25. W.-S. Ai, J. Liu, Q. Feng, P.-F. Zhai, P.-P. Hu, J. Zeng, S.-X. Zhang, Z.-Z. Li, L. Liu, X.-Y. Yan, and Y.-M. Sun, "Degradation of β-Ga2O3 Schottky barrier diode under swift heavy ion irradiation*," Chinese Phys. B 30(5), 056110 (2021).

26. S. Yadav, K. Das Gupta, and S. Dhar, "Controlling the persistence of photoconductivity through additional sub-bandgap photoexcitation in individual m-axial GaN nanowires," Journal of Applied Physics 126(4), 044302 (2019).

27. B.R. Tak, M.-M. Yang, Y.-H. Lai, Y.-H. Chu, M. Alexe, and R. Singh, "Photovoltaic and flexible deep ultraviolet wavelength detector based on novel β-Ga2O3/muscovite heteroepitaxy," Sci Rep 10(1), 16098 (2020).

28. B. Potì, A. Passaseo, M. Lomascolo, R. Cingolani, and M. De Vittorio, "Persistent photocurrent spectroscopy of GaN metal–semiconductor–metal photodetectors on long time scale," Applied Physics Letters 85(25), 6083–6085 (2004).

29. B.R. Tak, M.-M. Yang, M. Alexe, and R. Singh, "Deep-Level Traps Responsible for Persistent Photocurrent in Pulsed-Laser-Deposited β-Ga2O3 Thin Films," Crystals 11(9), 1046 (2021).